\def\isarxiv{1} 

\ifdefined\isarxiv
\documentclass[11pt]{article}

\usepackage[numbers]{natbib}

\else

\documentclass{article} 
\usepackage{microtype} 
\usepackage{graphicx}
\usepackage{subfig}
\usepackage{hyperref}
\usepackage{icml2024}

\fi

\usepackage{enumitem}
\usepackage{amsmath}
\usepackage{amsthm}
\usepackage{amssymb}
\usepackage{algorithm}
\usepackage{algpseudocode}
\usepackage{grffile}
\usepackage{wrapfig,epsfig}
\usepackage{url}
\usepackage{xcolor}
\usepackage{epstopdf}

\usepackage{bbm}
\usepackage{dsfont}

\allowdisplaybreaks

\ifdefined\isarxiv

\usepackage{tikz}
\usepackage{hyperref}  
\hypersetup{colorlinks=true,citecolor=blue,linkcolor=blue} 
\usetikzlibrary{arrows}
\usepackage[margin=1in]{geometry}

\else

\usepackage{microtype}
\usepackage{hyperref}
\definecolor{mydarkblue}{rgb}{0,0.08,0.45}
\hypersetup{colorlinks=true, citecolor=mydarkblue,linkcolor=mydarkblue}

\fi

\theoremstyle{plain}
\newtheorem{theorem}{Theorem}[section]
\newtheorem{lemma}[theorem]{Lemma}
\newtheorem{definition}[theorem]{Definition}

\newtheorem{remark}[theorem]{Remark}

\newcommand{\wt}{\widetilde}

\newcommand{\R}{\mathbb{R}}

\renewcommand{\tilde}{\wt}

\newcommand{\opt}{\mathsf{opt}}
\newcommand{\alg}{\mathsf{alg}}

\DeclareMathOperator{\poly}{poly}

\makeatletter
\newcommand*{\RN}[1]{\expandafter\@slowromancap\romannumeral #1@}
\makeatother

\usepackage{lineno}

\ifdefined\isarxiv

\else

\usepackage[textsize=tiny]{todonotes}

\icmltitlerunning{Sublinear Time Algorithm for Online Weighted Bipartite Matching}

\fi

\begin{document}

\ifdefined\isarxiv

\else 
\fi

\ifdefined\isarxiv
\title{Sublinear Time Algorithm for Online Weighted Bipartite Matching}

\date{}

\author{
Hang Hu\thanks{\texttt{h-hu19@mails.tsinghua.edu.cn}. Tsinghua University.}
\and 
Zhao Song\thanks{\texttt{zsong@adobe.com}. Adobe Research.}
\and 
Runzhou Tao\thanks{\texttt{runzhou.tao@columbia.edu}. Columbia University.}
\and 
Zhaozhuo Xu\thanks{\texttt{zx22@rice.edu}. Rice University.}
\and
Junze Yin\thanks{\texttt{junze@bu.edu}. Boston University.}
\and
Danyang Zhuo\thanks{\texttt{danyang@cs.duke.edu}. Duke University.}
}

\else

\twocolumn[
\icmltitle{Sublinear Time Algorithm for Online Weighted Bipartite Matching}



\icmlsetsymbol{equal}{*}

\begin{icmlauthorlist}
\icmlauthor{Firstname1 Lastname1}{equal,yyy}
\icmlauthor{Firstname2 Lastname2}{equal,yyy,comp}
\icmlauthor{Firstname3 Lastname3}{comp}
\icmlauthor{Firstname4 Lastname4}{sch}
\icmlauthor{Firstname5 Lastname5}{yyy}
\icmlauthor{Firstname6 Lastname6}{sch,yyy,comp}
\icmlauthor{Firstname7 Lastname7}{comp}
\icmlauthor{Firstname8 Lastname8}{sch}
\icmlauthor{Firstname8 Lastname8}{yyy,comp}
\end{icmlauthorlist}

\icmlaffiliation{yyy}{Department of XXX, University of YYY, Location, Country}
\icmlaffiliation{comp}{Company Name, Location, Country}
\icmlaffiliation{sch}{School of ZZZ, Institute of WWW, Location, Country}

\icmlcorrespondingauthor{Firstname1 Lastname1}{first1.last1@xxx.edu}
\icmlcorrespondingauthor{Firstname2 Lastname2}{first2.last2@www.uk}

\icmlkeywords{Machine Learning, ICML}

\vskip 0.3in
]



\printAffiliationsAndNotice{\icmlEqualContribution} 

\fi

\ifdefined\isarxiv
\begin{titlepage}
  \maketitle
  \begin{abstract}
Online bipartite matching is a fundamental problem in online algorithms. The goal is to match two sets of vertices to maximize the sum of the edge weights, where for one set of vertices, each vertex and its corresponding edge weights appear in a sequence. Currently, in the practical recommendation system or search engine, the weights are decided by the inner product between the deep representation of a user and the deep representation of an item. The standard online matching needs to pay $nd$ time to linear scan all the $n$ items, computing weight (assuming each representation vector has length $d$), and then deciding the matching based on the weights. However, in reality, the $n$ could be very large, e.g. in online e-commerce platforms. Thus, improving the time of computing weights is a problem of practical significance. In this work, we provide the theoretical foundation for computing the weights approximately. We show that, with our proposed randomized data structures, the weights can be computed in sublinear time while still preserving the competitive ratio of the matching algorithm.

  \end{abstract}
  \thispagestyle{empty}
\end{titlepage}


\else
\begin{abstract}
Online bipartite matching is a fundamental problem in online algorithms. The goal is to match two sets of vertices to maximize the sum of the edge weights, where for one set of vertices, each vertex and its corresponding edge weights appear in a sequence. Currently, in the practical recommendation system or search engine, the weights are decided by the inner product between the deep representation of a user and the deep representation of an item. The standard online matching needs to pay $nd$ time to linear scan all the $n$ items, computing weight (assuming each representation vector has length $d$), and then deciding the matching based on the weights. However, in reality, the $n$ could be very large, e.g. in online e-commerce platforms. Thus, improving the time of computing weights is a problem of practical significance. In this work, we provide the theoretical foundation for computing the weights approximately. We show that, with our proposed randomized data structures, the weights can be computed in sublinear time while still preserving the competitive ratio of the matching algorithm.
\end{abstract}

\fi

\section{Introduction}

Online bipartite matching is a fundamental problem, and researchers have spent decades to optimize its performance \cite{mehta2013online,feldman2009online2,karande2011online,devanur2011near,haeupler2011online,
manshadi2012online,jaillet2013online,agk+20,brubach2020attenuate,dil+21,huang2021online}.  
The high-level goal is simple: how to provide matching for two sets of vertices to maximize the weights on the edges when the vertices in one set appear sequentially. This theoretical problem attracts practitioners' attention because it is highly related to real-world recommendation or ranking systems, e.g., matching job applications to jobs~\cite{zjo16,lms+19}, 
matching users to suggested items in online shopping~\cite{jlk+15,swk+20},  
matching ads to websites for display~\cite{fgz+19}.

In these real-world scenarios, the weight of an edge (e.g., the probability that a user likes a suggested item) is not arbitrary. Instead, we can model each user and each item as a representation vector and assign the weight of an edge to be the inner product of the user's and the item's representation vectors. For example, in recommendation systems, it is very common to use matrix factorization/decomposition in order to get this representation vector from a giant user-item rating matrix~\cite{kbv09,hzk+16,xdz+17}. Next, a user's rating on a particular item is predicted by the similarity of the user representation and item representation. In other words, the efficiency of the online bipartite matching algorithm determines the inference latency of deep representation models in recommendation systems.

The existing online matching algorithms \cite{kvv90, fmmm09, aggarwal2011online, fhtz20} require a linear scan on the vertex set. For example, when there are $n$ items and users come sequentially, this means these algorithms need to pay $nd$ time (assuming $d$ is the length of feature vectors of items and users) just to calculate the weights for a user. In modern online shopping portals, $n$ can be very large (e.g., Amazon sells more than 12 million different items), and thus even algorithms linear in $n$ can be slow.

This raises an interesting \textit{theoretical} question:
\begin{center}
    {\it Can we design sublinear time online bipartite matching algorithms when the weight on an edge is the inner product of feature vectors on the vertices?}
\end{center}

This question is worthwhile to consider for two reasons. First, improving the performance of matching has real-world implications, especially in cases at a large scale (i.e., when $n$ is large). Second, we currently do not know any sublinear time algorithm for online bipartite matching, and we would like to show that sublinear time is feasible in a meaningful subset of the problem space, where edge weight is the inner product of vectors on the vertices.

We list our contributions as follows:
\begin{itemize}
\item Our first contribution is the first analysis of matching based on approximated weights. Prior analysis on matching algorithms depends on knowing the exact weights \cite{huang2019tight, HuangTWZ/FOCS/2020, ashlagi2019edge}. With prior analysis, it is unclear whether there is \textit{any} guarantee when weights are approximated. Our finding is that, surprisingly, existing matching algorithms are already robust to approximated weights. 
\item 
Our second contribution is that we design efficient data structures and algorithms to compute weights in sublinear time. Our data structures maintain succinct states about feature vectors of the $n$ items. For a user's feature vector, our data structures can calculate all the corresponding weights in $o(nd)$. 
\end{itemize}

\paragraph{Roadmap}

Section~\ref{sec:preli} introduces some related data structures and conclusions that will be utilized in our paper. 
Section~\ref{sec:matching} defines our research problem formally and studies some conclusions for matching.
Section~\ref{sec:data_structure} proposes some data structures useful for our algorithms. 
Section~\ref{sec:result} proposes several useful algorithms for our research problem.

\subsection{Our Results}

First, we define the online matching problem as follows.

\begin{definition}[Informal version of Definition \ref{def:main_problem}]
Consider a set of offline points $U\subset \mathbb{R}^d$, a set of online points $V\subset \mathbb{R}^d$ and a weight function $w:\mathbb{R}^d\times \mathbb{R}^d\to \mathbb{R}_{\ge 0}$ which gives a weight for each two 
positions in $\mathbb{R}^d$. Offline points are known initially but online points come only one at a time. We say an algorithm is an online matching algorithm if it can match every online point to an offline point without knowing the information of the online points coming after it, and make the total matching as large as possible. For convenience, we permit an offline point to match with multiple online points, but when computing the value of the matching, only the edge with maximal weight will be taken into account.

The goal of the online matching problem is to find an online matching algorithm that can find a matching as large as possible and has an update time as small as possible. 
\end{definition}

Then, we present our main results. Theorem \ref{thm:distance_weight_informal} shows the online matching problem where weight is defined as the distance between vertices, and obtains an algorithm with a competitive ratio approximately equal to $\frac{1}{2}$. Theorem \ref{thm:inner_product_weight_informal} shows the online matching problem where weight is defined as the inner product between vertices, and obtains an algorithm with a competitive ratio approximately equal to $\frac{1}{2}$. Theorem~\ref{thm:inner_product_weight_v2_informal} also shows the online matching problem with the inner product weight, but obtains a better result.

\begin{theorem}[Main result, informal version of Theorem \ref{thm:distance_weight}] \label{thm:distance_weight_informal}
Let $\epsilon \in (0,1), \delta \in (0,1)$. 

Then, there is an online bipartite matching algorithm (with $w(x,y)=\| x- y\|_2$) that takes 
\begin{align*}
    \tilde{O}(\epsilon^{-2}(n+d)\log(n/\delta))
\end{align*}
time for each coming vertex, and our algorithm satisfies 
\begin{align*}
    \alg \ge \frac{1}{2} (1-2\epsilon) \cdot \opt
\end{align*}
with probability at least $1-\delta$, where $n$ is the number of online points.
\end{theorem}

\begin{theorem}[Main result, informal version of Theorem \ref{thm:inner_product_weight}] \label{thm:inner_product_weight_informal}
Given $D>0, \epsilon\in(0,1), \delta\in(0,1)$, we suppose all the offline points $u$ satisfies $\|u\|_2\le D$. 

Then there is an algorithm of online bipartite matching (with $w(x,y)=\langle x, y\rangle$) that takes
\begin{align*}
    \tilde{O}(\epsilon^{-2}D^2(n+d)\log(n/\delta))
\end{align*}
time for each coming vertex, and our algorithm satisfies 
\begin{align*}
    \alg \ge \frac{1}{2}\opt-\frac{3}{2}n\epsilon
\end{align*}
with probability at least $1-\delta$, where $n$ is the number of online points.
\end{theorem}

\begin{theorem}[Main result, informal version of Theorem \ref{thm:inner_product_weight_v2}] \label{thm:inner_product_weight_v2_informal}
Given $D>0, \epsilon\in(0,1), \tau\in(0,1), \delta\in(0,1)$, we suppose that all the offline points $u$ satisfies $\|u\|_2\le D$.

Then there is an algorithm of online bipartite matching (with $w(x,y)=\langle x, y\rangle$) that takes
\begin{align*}
    O(dn^{f(1-\epsilon,\tau/D)+o(1)}\log(n/\delta))
\end{align*}
time (where $f(x,y)=\frac{1-y}{1-2xy+y}$) for each coming vertex, and our algorithm satisfies 
\begin{align*}
    \alg \ge \frac{1}{2}\min\{(1-\epsilon)\opt, \opt-n\tau\}
\end{align*}
with probability at least $1-\delta$, where $n$ is the number of online points.
\end{theorem}

\subsection{Related Work}

\paragraph{Online Weighted Bipartite Matching.} Online weighted bipartite matching is an important problem, and there has been plenty of work on this problem. Mehta~\cite{mehta2013online} provides a complete survey on this topic. One important opportunity to reduce the running time is to make assumptions on the arrival patterns~\cite{feldman2009online2,karande2011online,devanur2011near,haeupler2011online,
manshadi2012online,jaillet2013online,agk+20,brubach2020attenuate,dil+21,huang2021online}. This can usually lead to a better competitive ratio. Many researchers also explored more generalized settings~\cite{huang2018match, gamlath2019beating,gamlath2019online, huang2019tight, HuangTWZ/FOCS/2020, ashlagi2019edge}. Our work is different from theirs because we consider a specialized version of the online weighted bipartite matching problem, where the edge cost is the inner product of vectors on the vertices. We believe this specialized version is very important and interesting to study. We demonstrate that for this specialized version of the problem, we can even achieve sublinear time matching.

\paragraph{Data Structures for Machine Learning.} There has been a recent trend in the research community to innovate new data structures for efficient machine learning. For example, one line of work is to use approximated maximum inner product search data structures~\cite{cmfgts20,clp+21,dmzs21}. This can substantially reduce the overheads of backward propagation during training. Other data structure designs also exist including using LSH-based data structures~\cite{xcl+21}. Besides search data structure, randomized distance estimation data structures are also widely applied in the approximation of kernels~\cite{acw17,akk+20,swy+21} and attention matrices~\cite{wlk+20,cdw+21}. Our design is different from these works. We focus on the dynamic scenarios where the incremental insertion and deletion of vectors in the database should be allowed. As a result, we aim at the co-design of efficient data structure and dynamic mechanism for the overall running time speedups. 

\section{Preliminaries}\label{sec:preli}

This section is organized as follows: In section \ref{section:LSH}, we summarize some data structures related to locality-sensitive hashing. In section \ref{section:other}, we cite an asymmetric transformation and some results of the generalized submodular welfare maximization problem which are helpful for our latter proofs.

\paragraph{Notations.}
We use $\| \cdot \|_2$ to denote $\ell_2$ norm. For any function $f$, we use $\wt{O}(f)$ to denote $f \cdot \poly(\log f)$. For integer $n$, we use $[n]$ to denote $\{1, 2, \cdots, n \}$.  We use $\Pr[]$ to denote the probability. For a set $A$, we use $2^A$ to denote the power set of $A$, namely $2^A : = \{x \mid x \subseteq A\}$.

\subsection{Locality Sensitive Hashing}\label{section:LSH}
In this section, we introduce the standard definitions for locality sensitive hashing~(LSH) function family for approximate near neighbor search~\cite{im98,diim04,ainr14,ailrs15,ar15,iw18,alrw17,air18,dirw19,ccd+20,ll21}. We also include statements that provide the theoretical guarantee of using LSH for maximum inner product search.

\begin{definition}[Locality sensitive hashing, \cite{im98}]
Given parameter $R>0,c>1,0<p_2<p_1<1$, we say a family of hash function ${\cal H}$ is $(R,cR,p_1,p_2)$-sensitive if it satisfies the following conditions:
\begin{itemize}
    \item for any $x,y\in\mathbb{R}^d$, as long as $\|x-y\|_2\le R$, then 
    \begin{align*}
        \Pr_{h\in {\cal H}} [h(x)=h(y)]\ge p_1;
    \end{align*}
    \item for any $x,y\in\mathbb{R}^d$, as long as $\|x-y\|_2\ge cR$, then 
    \begin{align*}
        \Pr_{h\in {\cal H}} [h(x)=h(y)]\le p_2.
    \end{align*}
\end{itemize}
\end{definition}

\begin{definition}[Approximate Near Neighbor]

Let $c>1$ and $r\in (0,2)$. Given a
dataset $P\subset \mathbb{S}^{d-1}$ with $n$ points on the sphere, the $(c, r)$-Approximate Near Neighbor (ANN) problem aims at providing a data structure such that, for a query $q\in \mathbb{S}^{d-1}$ with the condition that there exists a vector $p$ in $P$ with $\|p-q\|_2\le r$, the data structure returns a vector $p'$ in $P$ within $\|p' - q\|_2\leq  c\cdot r$.
\end{definition}

\begin{definition}[Approximate Max-IP]

Let $c\in(0,1)$ and $\tau\in(0,1)$. Given a
dataset $P\subset \mathbb{S}^{d-1}$ with $n$ points on the sphere, the $(c,\tau)$-Maximum Inner Product Search (Max-IP) aims at providing a data structure such that, for a query $q\in \mathbb{S}^{d-1}$ with the condition that there exists a vector $p$ in $P$ with $\langle p,q \rangle \ge \tau$, the data structure returns a vector $p'$ in $P$ that $\langle p',q \rangle \ge c\cdot \tau$.
\end{definition}

\begin{theorem}[Andoni and Razenshteyn \cite{ar15}]\label{thm:ANN}
Let $c>1$ and $r\in(0,2)$ and 
\begin{align*}
    \rho=\frac{1}{2c^2-1}+o(1).
\end{align*}

Then, the $(c,r)$-ANN on a unit sphere $\mathbb{S}^{d-1}$ can be solved in space $O(n^{1+\rho}+dn)$ and query time $O(d\cdot n^{\rho})$.
\end{theorem}

\begin{theorem}[Theorem 8.2, page 19, \cite{ssx21}]\label{thm:max-IP}
Let $c\in(0,1)$ and $\tau\in(0,1)$. 

Given a dataset $Y\subset \mathbb{S}^{d-1}$ with $n$-points on the sphere, one can construct a data structure using ${\cal T}_{\mathrm{init}}$ preprocessing time and ${\cal S}_{\mathrm{space}}$ space so that for any query $x\in {\cal S}^{d-1}$, we take query time complexity $O(d\cdot n^{\rho}\log(1/\delta))$:
\begin{itemize}
    \item if $\mathsf{Max-IP}(x,Y) \ge \tau$, then we output a vector in $Y$ which is a $(c,\tau)-\mathsf{Max-IP}$ for $(x,Y)$ with failure probability at most $\delta$, where $\rho:=f(c,\tau)+o(1)$ and $\mathsf{Max-IP}(x, Y ) := \max_{y \in Y} \langle x, y\rangle$.
    \item otherwise, we output $\mathsf{fail}$.
\end{itemize}
Further, 
\begin{itemize}
    \item If 
       ${\cal T}_{\mathsf{init}}=O(dn^{1+\rho})\log(1/\delta)$
    and 
    \begin{align*}
        {\cal S}_{\mathsf{space}}=O(n^{1+\rho}+dn)\log(1/\delta),
    \end{align*}
    then we have
    \begin{align*}
        f(c,\tau)=\frac{1-\tau}{1-2c\tau+\tau}.
    \end{align*}
    \item If 
        ${\cal T}_{\mathsf{init}}=O(dn^{1+o(1)})\log(1/\delta)$
    and
    \begin{align*}
        {\cal S}_{\mathsf{space}}=O(n^{1+o(1)}+dn)\log(1/\delta),
    \end{align*}
    then we have
    \begin{align*}
        f(c,\tau)=\frac{2(1-\tau)^2}{(1-c\tau)^2}-\frac{(1-\tau)^4}{(1-c\tau)^4}.
    \end{align*}
\end{itemize}
\end{theorem}

\subsection{Other Useful Concepts} \label{section:other}
In this section, we provide useful concepts and tools for our theoretical analysis.

\begin{definition}[Submodular function \cite{s03}]
Let $U$ be a finite ground set. A function $f : 2^U \to \mathbb{R}$ is called submodular if for every $S \subset T \subset U$ and $u \in U \backslash T$, 
\begin{align*}
    f(S \cup \{u\}) - f(S) \ge f(T \cup \{u\}) - f(T).
\end{align*}
\end{definition}

We state a well-known transformation in literature \cite{ns15,xss21,ssx21,sxz22,sxyz22}.
\begin{definition}[Asymmetric transformation \cite{ns15}]
Let $Y\subseteq \mathbb{R}^d$ and $\|b\|_2\le 1$ for all $b\in B$. Let $a\in \mathbb{R}^d$ and $\|a\|_2\le C_a$. We define the following asymmetric transform:  
$
    P(b)=[
    b^{\top}, \sqrt{1-\|b\|_2^2}, 0 
    ]^{\top} $ and $
    Q(a)=[ 
    (aC_a^{-1})^{\top}, 0 , \sqrt{1-\|aC_a^{-1}\|_2^2} 
    ]^{\top}.
$

Therefore, we have 
\begin{itemize} 
 \item    $\|Q(a)-P(b)\|_2^2=2-2C_a^{-1}\langle a,b \rangle $, 
 \item $ \arg\max_{b\in B}\langle a,b \rangle=\arg\min_{b\in B}\|Q(a)-P(b)\|_2^2.
$
\end{itemize}
\end{definition}


\section{Online Weight Bipartite Matching With Approximate Weight}\label{sec:matching}

This section is organized as follows: In Section \ref{section:definition}, we give a formal definition of our research problem, the online matching problem with feature vector for each node. In Section \ref{section:1/2-approximation}, we show that there exists a simple greedy algorithm that can output a matching at least half of the optimal matching. In Section \ref{section:robustness}, we research the robustness of the algorithm in Section \ref{section:1/2-approximation}, and prove that it is robust to the error of input weights.

\subsection{Our Online Matching Problem}\label{section:definition}

We define our online matching problem as follows:
\begin{definition}[Online matching problem with feature vector for each node]\label{def:main_problem}
Given a weight function $w:\mathbb{R}^d\times\mathbb{R}^d \to \mathbb{R}_{\ge 0}$, we say an algorithm is an online matching algorithm, if it supports the following operations:
\begin{itemize}
    \item \textsc{Init}$(U=\{u_1,u_2,\cdots,u_n\}\subseteq \mathbb{R}^d)$. Given a finite point set $U\subseteq \mathbb{R}^d$, the algorithm preprocesses the set $U$.
    \item \textsc{Update}$(v\in \mathbb{R}^d)$. Given a point $v$, the algorithm chooses a $u\in U$ to match $v$.
    \item \textsc{Query}$()$. The algorithm outputs 
    \begin{align*} 
    \sum_{u\in U}\max_{v\in V, \text{$v$ is matched to $u$}}\{w(u,v)\},
    \end{align*}
    where $V$ is the set of all the given $v$ in \textsc{Update}.
\end{itemize}
We say an online matching algorithm is $\Gamma$-competitive (or $\Gamma$-approximate), if its \textsc{Query} can always output a value at least $\Gamma$ times of the maximal value among all the matching cases. The goal of this problem is to find the maximal $\Gamma$ such that a $\Gamma$-competitive online matching algorithm exists.
\end{definition}

We mainly study the following 2 versions of online matching problems, online matching problem with inner product weight ($w(u,v)=\langle u,v \rangle$) and online matching problem with distance weight ($w(u,v)=d(u,v)$). If we assume $u$ and $v$ are representations of their original features $u'$ and $v'$, the inner product weight corresponds to the linearized form of a kernel~\cite{rr07}. Moreover, the distance weight relates to metric learning~\cite{yj06,hoffer2015deep,xdz+17}.

\subsection{Greedy Algorithm is 1/2-approximation}\label{section:1/2-approximation}
Using the tool from previous work (Theorem~\ref{thm:approximation_ratio}), we can also show the following standard result,
\begin{theorem}[\cite{kvv90}]\label{thm:1/2-approximation}
For the online matching problem with a feature vector for each node, there is a $1/2$-approximate algorithm (see Algorithm \ref{alg:standard_greedy}).
\end{theorem}
\begin{proof}
Let $U$ denote the offline point set and $V$ denote the online point set and $|U|=n,|V|=m$. 

To transform the online matching problem into the setting of \cite{kvv90,bffg20}, let complete set ${\cal N}=U\times [m]$ and divide ${\cal N}$ into 
\begin{align*}
    P_1,P_2,\cdots,P_m,
\end{align*}
where  $P_i=U\times \{i\}$. 

Also, define matroid ${\cal M}$ such that its base consists of all the elements of $P_1\times P_2\times\cdots\times P_m$. Now we map every element $S\in {\cal M}$ to a matching of $U$ and $V$. 

Specifically, for each $i\in[m]$, 
\begin{itemize}
    \item if $|S\cap P_i|=1$, suppose $S\cap P_i=\{u_i\}$, we match $v_i$ to $u_i$;
    \item if $|S\cap P_i|=0$, we leave $v_i$ as unmatched. 
\end{itemize}

 For each $S\in {\cal M}$, let $M_S$ denote the matching mapped from $S$. Then we can define the matching function $f:{\cal M}\to \mathbb{R}_+$ by 
 \begin{align*}
     f(S)=|M_S|,
 \end{align*}
 where $|M_S|$ expresses the value of $M_S$.

Then Algorithm \ref{alg:standard_greedy} is $1/2$-approximation by  
Theorem \ref{thm:approximation_ratio} and the following Lemma \ref{thm:submodular}.
\end{proof}

\begin{lemma}[folklore, see \cite{kvv90} for an example]
\label{thm:submodular}
Matching function $f$ is non-negative submodular. 
\end{lemma}
\begin{proof}
Notice that for any $S\subset T\subset {\cal M}$ and $\{(u,i)\}\in {\cal M}$ such that 
\begin{align*}
    S\cup \{(u,i)\}\in {\cal M},T\cup \{(u,i)\}\in {\cal M},
\end{align*}
we have
\begin{align*}
    f(S\cup \{(u,i)\})-f(S)=\max\{w_{u,v_i}-\max_{e\in M_S, u\in e}\{w_e\},0\}
\end{align*} 
and 
\begin{align*}
    f(T\cup \{(u,i)\})-f(T)=\max\{w_{u,v_i}-\max_{e\in M_T, u\in e}\{w_e\},0\}.
\end{align*} 
Obviously $M_S\subseteq M_T$, so 
\begin{align*}
    \max_{e\in M_S, u\in e}\{w_e\}\le \max_{e\in M_T, u\in e}\{w_e\},
\end{align*} 
thus  
\begin{align*}
    f(S\cup \{(u,i)\})-f(S)\ge f(T\cup\{(u,i)\})-f(T),
\end{align*} 
hence $f$ is non-negative and submodular.
\end{proof}

\subsection{Approximate Weight Function Implies Slightly Worse Approximate Ratio}\label{section:robustness}
In this section, we research the cases when we only know the approximate weight function.

\begin{lemma} \label{lemma:main}
Let $\epsilon \in (0,0.1)$. 
Suppose that for any online node $v\in V$, for a general online matching problem $A=(U,V,w)$, we only know an approximated weight $\tilde{w}(u,v)$ of an edge $(u,v)$ rather than the real weight $w(u,v)$ where 
\begin{align*}
    (1-\epsilon)w(u,v)\le \tilde{w}(u,v)\le (1+\epsilon)w(u,v).
\end{align*}

Then there exists a greedy algorithm with competitive ratio $\frac{1}{2}(1-2\epsilon)$. 
\end{lemma}
\begin{proof}
Apply the online greedy algorithm stated in Theorem \ref{thm:1/2-approximation} on weight $\tilde{w}_{u,v}$. We prove that it is a $\frac{1}{2} (1-2\epsilon)$-approximation.

Suppose the original problem is $A=(U,V,w)$ with optimal matching $\opt$, and our algorithm gives a matching $\alg$. Define another online matching problem $A'=(U,V,\tilde{w})$ with optimal matching $\opt'$ and our algorithm gives a matching $\alg'$, since 
\begin{align*} 
(1-\epsilon)w_{u,v}\le \tilde{w}_{u,v}\le (1+\epsilon)w_{u,v},
\end{align*}
we have $\alg'\le (1+\epsilon)\alg$ and $\opt'\ge (1-\epsilon)\opt$.

Thus 
\begin{align*}
    \alg \ge & ~ \frac{1}{1+\epsilon}\alg' 
    \ge  ~ \frac{1}{2(1+\epsilon)}\opt' 
    \ge ~ \frac{1-\epsilon}{2(1+\epsilon)}\opt,
\end{align*} 
where the first step follows from $\alg' \le (1 + \epsilon) \alg$, the second step follows from Theorem \ref{thm:1/2-approximation}, and the last step follows from $\opt' \ge (1 - \epsilon) \opt$.

Thus, the conclusion follows from $\frac{1-\epsilon}{2(1+\epsilon)} \ge \frac{1}{2}(1 - \epsilon)^2 \ge \frac{1}{2}(1 - 2\epsilon)$. 
\end{proof}

\begin{lemma}\label{lemma:greedy_robustness}
Consider the general online matching problem $A=(U,V,w)$. 

Suppose that we have an oracle $P$ such that when every online node $v\in V$ comes if we define $w_u$ as the accumulated weight of $u$ and $\Delta w_u$ as the increments of the matching value after matching $v$ to $u$, $\forall u$, the oracle tells us a $u_0$ such that at least one of the following two inequalities holds: 
\begin{itemize}
    \item $\Delta w_{u_0}\ge (1-\epsilon) \cdot \max_{u \in [n]}\{\Delta w_u\}$  
    \item $\Delta w_{u_0}\ge \max_{u \in [n]}\{\Delta w_u\} - \tau$
\end{itemize}

Then there exists an algorithm that gives a matching $\alg$ such that 
\begin{align*} 
\alg \ge \frac{1}{2}\min\{(1-\epsilon) \opt, \opt-n\tau\}.
\end{align*}
\end{lemma}
\begin{proof}
Define the algorithm $\alg$: match $v$ to the $u_0$ given by the oracle $P$ whenever an online point $v$ comes. We prove the algorithm satisfies \begin{align*} 
\alg \ge \frac{1}{2}\min\{(1-\epsilon) \opt, \opt-n\tau\}.
\end{align*}

In fact, define another online matching problem $A'=(U,V,w')$: when an online point $v$ comes, suppose $P$ gives an oracle $u_0$, and for each $u\in U$, then we let $w'(u,v)=w_u+\Delta w_{u_0}$. That is, the definitions of $w'$ makes that: $\alg$ is a greedy algorithm of $A'$, 
\begin{align*}
    w'(u_0,v)=w(u_0,v)
\end{align*}
and for all $u$, if $\Delta w_{u_0}\ge (1-\epsilon)\max_{u \in [n]}\{\Delta w_u\}$, then
\begin{align*}
    w'(u,v) 
    = & ~ w_u+\Delta w_{u_0} \\
    \ge & ~ w_u+(1-\epsilon)\Delta w_{u} \\
    \ge & ~ (1-\epsilon)(w_u+\Delta w_u) \\
    = & ~ (1-\epsilon) w(u,v)
\end{align*}
where the first step follows from the definition of $w'(u, v)$, the second step follows from $\Delta w_{u_0} \ge (1-\epsilon)\Delta w_{u}$, the third step follows from $\epsilon \ge 0$, and the last step follows from the definition of $\Delta w_u$.

If $\Delta w_{u_0}\ge \max_{u \in [n]}\{\Delta w_u\} - \tau$, then 
\begin{align*}
    w'(u,v) = & ~  w_u+\Delta w_{u_0} \\
    \ge & ~ w_u+\Delta w_u-\tau \\
    = & ~ w(u,v)-\tau,
\end{align*}
where the first step follows from the definition of $w'(u, v)$, the second step follows from $\Delta w_{u_0}\ge \max_{u \in [n]}\{\Delta w_u\} - \tau$, and the last step follows from the definition of $\Delta w_u$.

Thus by Theorem \ref{thm:1/2-approximation},
\begin{align*}
    \alg 
    = & ~ \alg' \\
    \ge & ~ \frac{1}{2}\opt' \\
    \ge & ~ \frac{1}{2}\min\{(1-\epsilon) \opt, \opt-n\tau\},
\end{align*}
where the first step follows from the definition of $\alg'$, the second step follows from Theorem \ref{thm:1/2-approximation}, and the last step follows from for edge $e(u, v)$, $w'(u, v) \ge \frac{1}{2}\min\{(1-\epsilon) \opt, \opt-n\tau\}$.

This finishes the proof.
\end{proof}

\begin{lemma} \label{lemma:main'}
For general online matching problem $A=(U,V,w)$, suppose for any online node $v\in V$, we only know an approximated weight $\tilde{w}(u,v)$ of an edge $e_{u,v}$ rather than the real weight $w(u,v)$ where 
\begin{align*}
    w(u,v)-\epsilon\le \tilde{w}(u,v)\le w(u,v)+\epsilon.
\end{align*}

Then there exists a greedy algorithm that outputs $\alg$ such that 
\begin{align*}
    \alg \ge \frac{1}{2}\opt - \frac{3}{2}n\epsilon,
\end{align*}
where $n=|V|$ is the number of online points.
\end{lemma}
\begin{proof}
Apply the online greedy algorithm stated in Theorem \ref{thm:1/2-approximation} on weight $\tilde{w}_{u,v}$. We prove that it satisfies $\alg\ge \frac{1}{2}\opt - \frac{3}{2}n\epsilon$.

Suppose the original problem is $A=(U,V,w)$ with optimal matching $\opt$, and our algorithm gives a matching $\alg$. Define another online matching problem $A'=(U,V,\tilde{w})$ with optimal matching $\opt'$ and our algorithm gives a matching $\alg'$, since $w_{u,v}-\epsilon \le \tilde{w}_{u,v} \le w_{u,v}+\epsilon$, we have $\alg'\le \alg+n\epsilon$ and $\opt'\ge \opt-n\epsilon$, thus 
\begin{align*}
    \alg 
    \ge & ~ \alg'-n\epsilon \\
    \ge & ~ \frac{1}{2}\opt'-n\epsilon \\
    \ge & ~ \frac{1}{2}\opt-\frac{3}{2}n\epsilon,
\end{align*}
where the first step follows from $\alg' \le \alg + n\epsilon$, the second step follows from Theorem \ref{thm:1/2-approximation}, and the last step follows from $\opt' \ge \opt - n\epsilon$. 
\end{proof}

 
\section{Dynamic Data Structure}\label{sec:data_structure}

In this section, we present an inner product estimation data structure that approximately estimates the inner product between a query point $q$ and the given data points.
     
Due to the page limit, we include a detailed algorithm presentation in Appendix~\ref{sec:ADE}. The objective of the inner product estimation is to estimate the inner product of a given query with every vector in a dataset with improvements over the naive running time.

\begin{theorem}\label{thm:IPE}
There is a data structure that uses $\tilde{O}(\epsilon^{-2}D^2nd\log(1/\delta))$ space for the Online Approximate Adaptive Inner Product Estimation Problem with the following procedures:
\begin{itemize}
    \item \textsc{Init}$(  \{x_1, x_2, \dots, x_n\}\subset \R^d, \epsilon \in (0,1), \delta \in (0,1))$: Given data points $\{x_1, x_2, \dots, x_n\}\subset \R^d$ ($\| x_i \|_2 \leq D$,  
    for all $i \in [n]$), an accuracy parameter $\epsilon$ and a failure probability $\delta$ as input, the data structure preprocesses in time $\wt{O}(\epsilon^{-2} D^2 n d \log(1/\delta))$.
    \item \textsc{Update}$(i \in [n], z \in \R^d)$: Given index $i$ and coordinate $z$, the data structure replace $x_i$ by $z$ in time $\tilde{O}(\epsilon^{-2}D^2d\log(1/\delta))$.
    \item \textsc{Query}$(q \in \R^d)$: Given a query point $q \in \R^d$ (where $\| q \|_2 \leq 1$),  
    the \textsc{Query} operation takes $q$ as input and approximately estimates the inner product of $q$ and all the data points $\{x_1, x_2, \dots, x_n\}\subset \R^d$ in time $\tilde{O}(\epsilon^{-2}D^2(n+d)\log(1/\delta))$ i.e. it provides a set of estimates $\{\wt{w}_i\}_{i=1}^n$ such that:
    \begin{align*}
       \forall i \in[n], \langle q,x_{i}\rangle - \epsilon \leq \wt{w}_{i} \leq \langle q,x_{i}\rangle + \epsilon
    \end{align*}
     with probability at least $1 -\delta$, even for a sequence of adaptively chosen queries.
\end{itemize}
\end{theorem}

 
\section{Combination}\label{sec:result}

In Section \ref{section:distance_weight}, we utilize the adaptive distance estimation data structure proposed in \cite{cn22} to give an approximated algorithm with $\tilde{O}(n+d)$ update time for the online matching problem with distance weight. In Section \ref{section:inner_product_weight}, we utilize the inner product estimation data structure stated in Section \ref{sec:ADE} to give an approximated algorithm with $\tilde{O}(n+d)$ for the online matching problem with inner product weight. In Section \ref{section:inner_product_weight_v2}, we utilize the max-IP data structure proposed in \cite{ssx21} to give an approximated algorithm with sublinear update time for the online matching problem with inner product weight.

\subsection{Online Matching Problem with Distance Weight}\label{section:distance_weight}

The goal of this section is to prove Theorem \ref{thm:distance_weight}, which analyzes the online matching problem with distance weight.

\begin{theorem}\label{thm:distance_weight}
Consider the online bipartite matching problem with distance weight, for any $\epsilon\in(0,1),\delta\in(0,1)$, there exists a data structure that uses $\tilde{O}(\epsilon^{-2}nd\log(n/\delta))$ space that supports the following operations
\begin{itemize}
    \item \textsc{Init}$(\{x_1,x_2,\cdots,x_n\}\subseteq \mathbb{R}^d, \epsilon\in(0,1), \delta\in(0,1))$. It takes a series of offline points $x_1,x_2,\cdots,x_n$, a precision parameter $\epsilon$ and a failure tolerance $\delta$ as inputs and runs in  $\tilde{O}(\epsilon^{-2}nd\log(n/\delta))$ time
    \item \textsc{Update}$(y\in\mathbb{R}^d)$. It takes an online point $y$ as inputs and runs in $\tilde{O}(\epsilon^{-2}(n+d)\log(n/\delta))$, time.
    \item \textsc{Query}. It outputs the matching value between offline points and known online points in $O(1)$ time such that it has competitive ratio $\frac{1}{2}(1-2\epsilon)$ with probability at least $1-\delta$, where $n$ is the number of online points.
\end{itemize}
\end{theorem}

We prove the data structure (see Algorithm \ref{alg:distance_weight}) satisfies the requirements of Theorem \ref{thm:distance_weight} by proving the following lemmas.
\begin{lemma}\label{lemma:distance_weight_init}
The procedure \textsc{Init} (Algorithm \ref{alg:distance_weight}) in Theorem \ref{thm:distance_weight} runs in time $\tilde{O}(\epsilon^{-2}nd\log(n/\delta))$.
\end{lemma}
\begin{proof}
This is because the member Ade has \textsc{Init} time $\tilde{O}(\epsilon^{-2}nd\log(n/\delta))$.
\end{proof}

\begin{lemma}\label{lemma:distance_weight_update}
The procedure \textsc{Update} (Algorithm \ref{alg:distance_weight}) in Theorem \ref{thm:distance_weight} runs in time $\tilde{O}(\epsilon^{-2}(n+d)\log(n/\delta))$ and suppose the member $s$ maintains a $\frac{1}{2}(1-2\epsilon)$-approximate matching before running \textsc{Update}, then $s$ also maintains a $\frac{1}{2}(1-2\epsilon)$-approximate matching after running \textsc{Update} with probability at least $1-\delta$.
\end{lemma}
\begin{proof}

The running time of each line is stated as follows: 
\begin{itemize}
    \item Line \ref{line:dw_update_1} calls Ade.\textsc{Query}, taking $\tilde{O}(\epsilon^{-2}D^2(n+d)\log(n/\delta))$ time by Theorem \ref{thm:cn22}.
    \item Line \ref{line:dw_update_2} computes $i_0$, taking $O(n)$ time.
    \item Line \ref{line:dw_update_3} updates $u_{i_0}$, taking at most $\tilde{O}(n)$ time.
    \item Line \ref{line:dw_update_4} updates $s$, taking $O(1)$ time.
    \item Line \ref{line:dw_update_5} updates $w_{i_0}$, taking $O(1)$ time.
\end{itemize}

Thus procedure \textsc{Update} runs in $\tilde{O}(\epsilon^{-2}D^2(n+d)\log(n/\delta))$ time.

For correctness, suppose $s$ maintains a $\frac{1}{2}(1-2\epsilon)$-approximate matching, then as long as Ade.\textsc{Update} succeeds, $s$ still maintains a $\frac{1}{2}(1-2\epsilon)$-approximate matching according to Theorem \ref{thm:1/2-approximation} and Lemma \ref{lemma:main}. And since the probability of Ade.\textsc{Update} succeeds is at least $1-\delta$ according to Theorem \ref{thm:cn22}, we have $s$ still maintains a $\frac{1}{2}(1-2\epsilon)$-approximate matching with probability at least $1-\delta$.
\end{proof}

\begin{lemma}\label{lemma:distance_weight_query}
The procedure \textsc{Query} (Algorithm \ref{alg:distance_weight}) in Theorem \ref{thm:distance_weight} correctly outputs a $\frac{1}{2}(1-2\epsilon)$-approximate matching with probability at least $1-\delta$, where $n$ is the number of online points.
\end{lemma}
\begin{proof}
Since in each time, \textsc{Update} succeeds with probability at least $1-\delta/n$, by union bound, the probability of success is at least $1-\delta$.
\end{proof}

\subsection{Online Matching Problem with Inner Product Weight}\label{section:inner_product_weight}

The goal of this section is to prove Theorem \ref{thm:inner_product_weight}, which analyzes the online bipartite matching problem with inner product weight.

\begin{theorem}\label{thm:inner_product_weight}
Consider the online bipartite matching problem with the inner product weight, for any $\epsilon\in(0,1),\delta\in(0,1)$ there exists a data structure that uses $\tilde{O}(\epsilon^{-2}D^2nd\log(n/\delta))$ space that supports the following operations
\begin{itemize}
    \item \textsc{Init}$(\{x_1,x_2,\cdots,x_n\}\subseteq \mathbb{R}^d, \epsilon\in(0,1), \delta\in(0,1))$. Given a series of offline points $x_1,x_2,\cdots,x_n$ ($\|x_i\|_2\le D$, for all $i\in[n]$), a precision parameter $\epsilon$ and a failure tolerance $\delta$ as inputs, this data structure preprocesses in $\tilde{O}(\epsilon^{-2}D^2nd\log(n/\delta))$ time. 
    \item \textsc{Update}$(y\in\mathbb{R}^d)$. It takes an online point $y$ as inputs and runs in $\tilde{O}(\epsilon^{-2}D^2(n+d)\log(n/\delta))$ time.
    \item \textsc{Query}. It outputs the matching value between offline points and known online points in $O(1)$ time such that $\alg \ge \frac{1}{2}\opt - \frac{3}{2}n\epsilon$ with probability at least $1-\delta$, where $n$ is the number of online points.
\end{itemize}
\end{theorem}

We prove the data structure (see Algorithm \ref{alg:inner_product_weight}) satisfies the requirements of Theorem \ref{thm:inner_product_weight} by proving the following lemmas.

\begin{lemma}\label{lemma:inner_product_weight_init}
The procedure \textsc{Init} in Theorem \ref{thm:inner_product_weight} (Algorithm \ref{alg:inner_product_weight}) runs in time $\tilde{O}(\epsilon^{-2}D^2nd\log(n/\delta))$.
\end{lemma}
\begin{proof}
This is because the member Ipe has \textsc{Init} time $\tilde{O}(\epsilon^{-2}D^2nd\log(n/\delta))$.
\end{proof}

\begin{lemma}\label{lemma:inner_product_weight_update}
The procedure \textsc{Update} in Theorem \ref{thm:inner_product_weight} (Algorithm \ref{alg:inner_product_weight}) runs in time $\tilde{O}(\epsilon^{-2}D^2(n+d)\log(n/\delta))$.
\end{lemma}

\begin{lemma}\label{lemma:inner_product_weight_query}
The procedure \textsc{Query} in Theorem \ref{thm:inner_product_weight} (Algorithm \ref{alg:inner_product_weight}) correctly outputs a matching $\alg$ such that $\alg \ge \frac{1}{2}\opt - \frac{3}{2}n\epsilon$ with probability at least $1-\delta$, where $n$ is the number of online points.
\end{lemma}
We defer the proof of the above two Lemmas into Appendix~\ref{sec:app_ip}.

\subsection{Better Result for Online Matching Problem with Inner Product Weight}\label{section:inner_product_weight_v2}

The goal of this section is to prove Theorem \ref{thm:inner_product_weight_v2}, which analyzes the online bipartite matching problem with inner product weight. Compared with Theorem \ref{thm:inner_product_weight}, Theorem \ref{thm:inner_product_weight_v2} has higher efficiency. We delay the proofs into Appendix~\ref{sec:app_ip}.

\begin{theorem}\label{thm:inner_product_weight_v2}
Consider the online bipartite matching problem with the inner product weight, for any $\epsilon\in(0,1),\tau\in(0,1),\delta\in(0,1)$, there exists a data structure that uses 
    $O((n^{1+\rho}+dn)\log(n/\delta))$
space, where 
\begin{align*}
    \rho=f(1-\epsilon,\tau/D)+o(1)
\end{align*}
and
\begin{align*}
    f(x,y):=\frac{1-y}{1-2xy+y}
\end{align*}
that supports the following operations
\begin{itemize}
    \item \textsc{Init}$(\{x_1,x_2,\cdots,x_n\}\subseteq \mathbb{R}^d, \epsilon\in(0,1), \tau\in(0,1), \delta\in(0,1))$. Given a series of offline points $x_1,x_2,\cdots,x_n$ ($\|x_i\|_2\le D$, for all $i\in[n]$), precision parameters $\epsilon$, $\tau$ and a failure tolerance $\delta$ as inputs, this data structure preprocesses in $O(dn^{1+\rho}\log(n/\delta))$ time  
    .
    \item \textsc{Update}$(y\in\mathbb{R}^d)$. It takes an online point $y$ as inputs $\|y\|_2\le 1$ and runs in $O(dn^{\rho}\log(n/\delta))$  
    time.
    \item \textsc{Query}. It outputs the matching value between offline points and known online points in $O(1)$ time such that 
    \begin{align*}
        \alg \ge \frac{1}{2}\min\{(1-\epsilon) \opt, \opt-n\tau\}
    \end{align*}
    with probability at least $1-\delta$,  
    where $n$ is the number of online points.
\end{itemize}
\end{theorem}
\begin{remark}
Intuitively, Theorem~\ref{thm:inner_product_weight_v2} is saying if $\epsilon$ is smaller, then the exponent of $n$ is larger, and if $\tau$ is smaller, then the exponent of $n$ is larger. From the lower bound on $\alg$, we also know that if $\epsilon$ is smaller and $\tau$ is smaller, then the algorithm's quality is better. Thus, if $\epsilon$ and $\tau$ are smaller, then the running time is larger, and the algorithm's quality is better. If $\epsilon$ and $\tau$  are larger, then the running time is smaller, and the algorithm's quality is worse. We can regard $\epsilon$ and $\tau$ as trade-off parameters from running and quality of algorithms.

For simplicity, let us consider $D=1$ and assume that $\opt \geq n$. Then the additive error $\opt-n \cdot \tau$ becomes a relative error which is $(1-\tau) \cdot \opt$. The $\alg$ is at least $0.5 \cdot \min \{(1-\epsilon)\opt, (1-\tau)\opt\}$. So in this case, it's natural to choose $\epsilon= \tau$. We compute several examples in this case below for reference.

Case 1. Choose $\epsilon = \tau = 0.75$, in this algorithm's performance is greater than 
 $\frac{1}{8} \cdot \opt$
 and the time is 
 \begin{align*}
    d \cdot n^{f(1-\epsilon, \tau)} = d \cdot n^{f(0.25, 0.75)} = d \cdot n^{0.181818}.
\end{align*}

Case 2. Choose $\epsilon = \tau = 0.5$, in this algorithm's performance is greater than 
 $\frac{1}{4} \cdot \opt$
 and the time is 
\begin{align*}
    d \cdot n^{f(1-\epsilon,\tau)} = d \cdot n^{f(0.5,0.5)} = d \cdot n^{0.5}.
\end{align*}

Case 3. Choose $\epsilon = \tau = 0.25$, in this algorithm's performance is greater than 
 $\frac{3}{8} \cdot \opt$
 and the time is 
 \begin{align*}
     d \cdot n^{f(1-\epsilon,\tau)} = d \cdot n^{f(0.75,0.75)} = d \cdot n^{0.857143}
 \end{align*}

The above three cases show that the algorithm performance is better. Meanwhile, the running time is slower.
\end{remark}

\section{Conclusion}\label{sec:conclusion}

Online bipartite matching is a well-known problem and has attracted numerous researchers to work on it. Our observation is that in many scenarios, the weight on an edge is not arbitrary. Instead, the weight of an edge is an inner product of feature vectors on the vertices. This creates a new opportunity to design a sublinear time algorithm for online bipartite matching. To match a set of incoming users to $n$ items, if the feature vector on a user and an item has a dimension of $d$, a standard online matching algorithm has to spend $O(nd)$ just to compute all the inner products. We provide the first algorithm that can approximate this computation in $o(nd)$ while maintaining a state-of-the-art competitive ratio. Our result is mainly an upper bound, we believe proving a lower bound is a very interesting future direction. 

\section*{Impact Statement} 

We use our algorithm to reduce the need for repetitive experimentation for different fields. We achieve this through strong theoretical foundations. Though it is inevitable to lead to more carbon emissions while implementing the algorithm, we hold that the potential benefits to the environment are far greater than its disadvantages because it can reduce the number of experiments. 

\ifdefined\isarxiv
\bibliographystyle{alpha}
\bibliography{ref}
\else
\bibliographystyle{icml2024}
\bibliography{ref}
\fi

\clearpage
\newpage
\onecolumn
\appendix 
\section*{Appendix}

\paragraph{Roadmap}
We present a roadmap here for better illustration. 
Section~\ref{sec:app_preli}, we state some basic notations and well-known tools from literature. 
Section \ref{sec:missing_proofs} presents our 
inner product estimation algorithm and its complete proof. 
Section \ref{sec:app_alg} gives a detailed version of the online matching algorithm used in our paper. 
Section \ref{sec:missing_details} presents the explanation of our algorithm. Section~\ref{sec:app_ip} provides missing proofs for inner product weight setting.

\section{Preliminaries}\label{sec:app_preli}
\subsection{Notations}
For a vector $u$, we use $\| u \|_2$ to denote the entry-wise $\ell_2$ norm of $u$. For two vectors $u,v$, we use $\langle u,v \rangle$ to denote their inner product. For any function $f$, we use $\wt{O}(f)$ to denote $O(f\cdot \poly(\log f))$. For a vector $u$ or a matrix $A$, we use $u^\top$ or $A^\top$ to denote its transpose. For any positive integer, we use $[n]$ to denote the set $\{1,2,\cdots,n\}$.

\subsection{Backgrounds on Submodular}

\begin{definition}[Generalized Submodular Welfare Maximization problem, page 1, \cite{bffg20}]
A ground set ${\cal N}$ is partitioned into disjoint non-empty sets $P_1,P_2,\cdots,P_m$. The goal is to choose a subset $S\subseteq{\cal N}$ that contains at most one element from each set $P_i$ and maximizes a given non-negative monotone submodular function $f$.
\end{definition}

 We state a well-known tool from previous work \cite{kvv90} (see Theorem 1 in \cite{bffg20} as an example, see Theorem 1 in lecture notes as another example\footnote{\url{https://www.cis.upenn.edu/~aaroth/courses/slides/privacymd/Lecture7.pdf}}).
\begin{theorem}[\cite{kvv90}]
\label{thm:approximation_ratio}
For generalized submodular welfare maximization problem, greedy algorithm \ref{alg:bffg} achieves an approximation of exactly 1/2.
\end{theorem}

\begin{algorithm}\caption{Greedy}\label{alg:bffg}
\begin{algorithmic}[1]
\State Initialize: $A_0\gets \Phi$
\For {$i \in [m]$}
    \State Let $u_i$ be the element $u\in P_i$ maximizing $f(u~|~A_{i-1}) := f(A_{i-1}\cup \{u\}) - f(A_{i-1})$.  
    \State $A_i\gets A_{i-1}\cup \{u_i\}$
\EndFor
\end{algorithmic}
\end{algorithm}

\section{Inner Product Estimation Data Structure} \label{sec:missing_proofs}

In this section, we provide a detailed version of the inner product estimation data structure used in our paper. We start with the formal description in Section~\ref{sec:ADE}, which includes the algorithm and its theoretical guarantee. Next, Section~\ref{sec:cn22} introduces the dynamic distance estimation data structure that used as a building block for our algorithm. Finally, we include the complete proof of our theoretical result in Section~\ref{sec:proof_of_IPE}.

\begin{algorithm}[!ht] \caption{Dynamic inner product estimation}\label{alg:inner_product_estimation}
\begin{algorithmic}[1]
\State {\bf data structure} \textsc{DynamicInnerProductEstimation} \Comment{Theorem \ref{thm:IPE}}
\State {\bf members}
\State \hspace{4mm} ADE Ade \Comment{Theorem \ref{thm:cn22}}
\State {\bf end members}
\State 
\Procedure{Init}{$x_1,x_2,\cdots,x_n,\epsilon,\delta$} \Comment{Lemma \ref{lemma:IPE_init} }  
    \State $\epsilon_0 \gets \frac{2\epsilon}{3D}$
    \State Ade.\textsc{Init}$(Q(x_1),Q(x_2),\cdots,Q(x_n),\epsilon_0,\delta)$
\EndProcedure 
\State 
\Procedure{Update}{$i,z$} \Comment{Lemma \ref{lemma:IPE_update} }  
    \State Ade.\textsc{Update}$(i,Q(z))$
\EndProcedure 
\State 
\Procedure{Query}{$q$} \Comment{Lemma \ref{lemma:IPE_query} }  
    \State $\wt{d}_1, \wt{d}_2, \cdots, \wt{d}_n \gets$  Ade.\textsc{Query}$(P(q))$
    \For {$i=1,2,\cdots,n$}
        \State $\wt{w}_i \gets 1-\frac{1}{2}\wt{d}_i^2$
    \EndFor 
    \State \Return $\{\wt{w}_i\}_{i=1}^n$
\EndProcedure 
\end{algorithmic}
\end{algorithm}

\subsection{Formal Description} \label{sec:ADE}

In this section, we present the theorem  and the corresponding algorithm. The algorithm is shown in Algorithm~\ref{alg:inner_product_estimation}. Along with the algorithm, we present the theoretical guarantees as below:

\begin{theorem}[Formal version of Theorem \ref{thm:IPE}] \label{thm:IPE_formal}
For the Online Approximate Adaptive Inner Product Estimation Problem, we solve it via
data structure that requires  $\tilde{O}(\epsilon^{-2}D^2nd\log(1/\delta))$ space. Moreover, the data structure supports three procedures:
 
\begin{itemize}
    \item \textsc{Init}$(  \{x_1, x_2, \dots, x_n\}\subset \R^d, \epsilon \in (0,1), \delta \in (0,1))$: 
    Given a dataset $\{x_1, x_2, \dots, x_n\}\subset \R^d$ ($\| x_i \| \leq D$,
    for all $i \in [n]$), the data structure preprocess it in time $\wt{O}(\epsilon^{-2} D^2 n d \log(1/\delta))$, where $\epsilon$ is the error parameter and $\delta$ is the failure probability.
 
    \item \textsc{Update}$(i \in [n], z \in \R^d)$: The data structure update $x_i$ with $z$ using $\tilde{O}(\epsilon^{-2}D^2d\log(1/\delta))$ time.
 
    \item \textsc{Query}$(q \in \R^d)$:
    The data structure estimates the inner product of a query $q$ with all data points $\{x_1, x_2, \dots, x_n\}\subset \R^d$ in time $\tilde{O}(\epsilon^{-2}D^2(n+d)\log(1/\delta))$. In other words, the data structure provides $\{\wt{w}_i\}_{i=1}^n$ that
    \begin{align*}
       \forall i \in[n], \langle q,x_{i}\rangle - \epsilon \leq \wt{w}_{i} \leq \langle q,x_{i}\rangle + \epsilon
    \end{align*}
     with failure probability at most $\delta$, even for a sequence of adaptively chosen queries.
\end{itemize}
\end{theorem}

In this next following sections, we aim at proving our statement.

\subsection{Auxiliary Data Structure: Dynamic Distance Estimation} \label{sec:cn22}

Before formally proving Theorem \ref{thm:cn22}, we first present a distance estimation data structure (ADE) from previous work.

\begin{theorem}[Theorem 1.4 in \cite{cn22}]\label{thm:cn22}
Let $\epsilon \in (0,0.1)$ denote an accuracy parameter. Let $\delta$ denote the failure probability. 
There is a data structure for the Online Approximate Adaptive Distance Estimation Problem uses $\tilde{O}(\epsilon^{-2}nd\log(1/\delta))$ space with the following procedures:
\begin{itemize}
    \item \textsc{Init}$(  \{x_1, x_2, \dots, x_n\}\subset \R^d, \epsilon \in (0,1), \delta \in (0,1))$: Given data points $\{x_1, x_2, \dots, x_n\}\subset \R^d$, an accuracy parameter $\epsilon$ and a failure probability $\delta$ as input, the data structure preprocesses in time $\tilde{O}(\epsilon^{-2}nd\log(1/\delta))$.
    \item \textsc{Update}$(i \in [n], z \in \R^{d})$.  
    This procedure takes $i$ and $z$ as inputs and replace $x_i$ by $z$. It takes $\tilde{O}(\epsilon^{-2}d\log(1/\delta))$ time.  
    \item \textsc{Query}$(q \in \R^d)$: Given a query point $q \in \R^d$, the \textsc{Query} operation takes $q$ as input and approximately estimates the Euclidean distances from $q$ to all the data points $\{x_1, x_2, \dots, x_n\}\subset \R^d$ in time $\tilde{O}(\epsilon^{-2}(n+d)\log(1/\delta))$ i.e. it provides a set of estimates $\{\wt{d}_i\}_{i=1}^n$ such that:
    \begin{align*}
       \forall i \in[n], (1-\epsilon)\|q-x_{i}\|_{2} \leq \wt{d}_{i} \leq(1+\epsilon)\|q-x_{i}\|_{2} 
    \end{align*}
     with probability at least $1 -\delta$, even for a sequence of adaptively chosen queries.
\end{itemize}
\end{theorem}

\subsection{Proof of Our Results} \label{sec:proof_of_IPE}

In this section, we give the proof of Theorem \ref{thm:IPE_formal}.

\begin{proof} 
We prove Theorem \ref{thm:IPE_formal} by proving the following Lemma \ref{lemma:IPE_init}, which focuses on procedure \textsc{Init}, Lemma \ref{lemma:IPE_update}, which focuses on procedure \textsc{Update} and Lemma  \ref{lemma:IPE_query}, which focuses on procedure \textsc{Query}.
\end{proof}
\begin{lemma}\label{lemma:IPE_init}
The procedure \textsc{Init} in  
Algorithm \ref{alg:inner_product_estimation} runs in time $\tilde{O}(\epsilon^{-2}D^2nd\log(1/\delta))$.
\end{lemma}
\begin{proof}
This is because the member Ade has \textsc{Init} time 
\begin{align*}
\tilde{O}(\epsilon_0^{-2}nd\log(1/\delta))=\tilde{O}(\epsilon^{-2}D^2nd\log(1/\delta))
\end{align*}
according to Theorem \ref{thm:cn22}.
\end{proof}

\begin{lemma}\label{lemma:IPE_update}
The procedure \textsc{Update} in Algorithm \ref{alg:inner_product_estimation} runs in time $\tilde{O}(\epsilon^{-2}D^2d\log(1/\delta))$.
\end{lemma}
\begin{proof}
This is because the member Ade has \textsc{Update} time \begin{align*}
\tilde{O}(\epsilon_0^{-2}d\log(1/\delta))=\tilde{O}(\epsilon^{-2}D^2d\log(1/\delta))
\end{align*}

according to Theorem \ref{thm:cn22}.
\end{proof}

\begin{lemma}\label{lemma:IPE_query}
The procedure \textsc{Query} in Algorithm \ref{alg:inner_product_estimation} outputs $\{\wt{w}_i\}_{i=1}^n$ correctly and runs in time $\tilde{O}(\epsilon^{-2}D^2(n+d)\log(1/\delta))$.
\end{lemma}
\begin{proof}
For the correctness of procedure \textsc{Query}, denote $\|Q(x_i)-P(y)\|_2$ by $d_i$, using asymmetric transformation we have
\begin{align*}
    \langle x_i,y \rangle &= D - \frac{D}{2}d_i^2,
\end{align*}
and according to Theorem \ref{thm:cn22}, Ade.\textsc{Query}($P(y)$) gives a series of   
$\{\wt{d}_i\}_{i=1}^n$ with $(1-\epsilon_0)d_i \le \wt{d}_i \le (1+\epsilon_0)d_i$, thus 
\begin{align*}
    \wt{w}_i = & ~ D-\frac{D}{2}\wt{d}_i^2 \\
    \le & ~ D-\frac{D}{2}(1-\epsilon_0)^2d_i^2 \\
    = & ~ \langle x_i,y \rangle + \frac{D}{2}(2\epsilon_0-\epsilon_0^2)d_i^2 \\
    \le & ~ \langle x_i,y \rangle + \epsilon,
\end{align*}
where the second step follows from $ \wt{d}_i\geq (1-\epsilon_0)d_i$, the third step follows from $\langle x_i, y \rangle = D -\frac{D}{2}d_i^2$ and the last step follows from  $\frac{D}{2}(2\epsilon_0-\epsilon_0^2)\le D\epsilon_0 \le \epsilon$.  

Similarly, we can prove the other direction:
\begin{align*}
    \wt{w}_i = & ~  D-\frac{D}{2}\wt{d}_i^2 \\
    \ge & ~ D-\frac{D}{2}(1+\epsilon_0)^2d_i^2 \\
    \ge & ~ \langle x_i,y \rangle - \frac{D}{2}(2\epsilon_0+\epsilon_0^2)d_i^2 \\
    \ge & ~ \langle x_i,y \rangle - \epsilon,
\end{align*}
where the second step follows from  $ \wt{d}_i\leq (1+\epsilon_0)d_i$, the third step follows from $\langle x_i, y \rangle = D - \frac{D}{2} d_i^2$, and the last step follows from $\frac{D}{2}(2\epsilon_0+\epsilon_0^2)\le \frac{3}{2}D\epsilon_0 \le \epsilon$.

And for the running time, it holds because the member Ade has \textsc{Query} time 
\begin{align*} 
\tilde{O}(\epsilon_0^{-2}(n+d)\log(1/\delta))=\tilde{O}(\epsilon^{-2}D^2(n+d)\log(1/\delta))
\end{align*}
according to Theorem \ref{thm:cn22}.
\end{proof}

\section{Our Algorithms} \label{sec:app_alg}

In this section, we present our algorithm for our online matching problem with distance weight. Note that
Section \ref{sec:result} has shown the three results which all give a solution to online matching problems with certain weights. Here we give the specific algorithm.

Firstly, we present Algorithm \ref{alg:distance_weight}, which corresponds to Theorem \ref{thm:distance_weight}. Here the weight is described by the distance. We presents a data structure which contains: (1) a series of $d$-dimensional vectors $x_1,x_2,\cdots,x_n$ used to record the coordinates of each offline point, (2) a series of auxiliary data structures $u_1,u_2,\cdots,u_n$ where each $u_i$ records the information of online points which are matched to $x_i$, (3) a series of real numbers $w_1,w_2,\cdots,w_n$ where each $w_i$ records the current maximal edge weight of $x_i$, (4) a online point matched to $x_i$, (5) a real number $s$ which records the current matching value, (6) an auxiliary adaptive distance estimation data structure ADE.

\begin{algorithm}[!ht] \caption{Algorithm for online matching problem with distance weight}\label{alg:distance_weight}
\begin{algorithmic}[1]
\State {\bf data structure} \textsc{DistanceMatching} \Comment{Theorem \ref{thm:distance_weight} }  
\State {\bf members}
\State \hspace{4mm} $x_1, x_2, \cdots x_n \in \R^{d}$ 
\State \hspace{4mm} $u_1,u_2,\cdots,u_n$ (linklist or tree)
\State \hspace{4mm} $w_1,w_2,\cdots,w_n$ (accumulated weights of each offline point)
\State \hspace{4mm} $s$ (matching value)
\State \hspace{4mm} \textsc{ADE} $\mathrm{Ade}$
\State {\bf end members}
\State 
\Procedure{Init}{$x_1, \cdots, x_n, \epsilon, \delta$} \Comment{Lemma \ref{lemma:distance_weight_init} }  
    \For {$i=1,2,\cdots,n$}
        \State $x_i \gets x_i$
        \State Init $u_i$
        \State $w_i \gets 0$
    \EndFor 
    \State $s \gets 0$
    \State $\mathrm{Ade}$.\textsc{Init}($x_1, \cdots, x_n, \epsilon, \delta / n$)
\EndProcedure 
\State 
\Procedure{Update}{$y \in \R^d$} \Comment{Lemma \ref{lemma:distance_weight_update} }  
    \State $\wt{d}_1,\wt{d}_2,\cdots,\wt{d}_n \gets \mathrm{Ade}.\textsc{Query}(y)$ \label{line:dw_update_1}
    \State $i_0 \gets \arg\max_{i}\{\wt{d}_i-w_i\}$ \label{line:dw_update_2}
    \State $u_{i_0}.\textsc{Insert}(y)$ \label{line:dw_update_3}
    \State $s \gets s + \max\{0,\wt{d}_{i_0}-w_{i_0}\}$ \label{line:dw_update_4}
    \State $w_{i_0} \gets \max\{w_{i_0},\wt{d}_{i_0}\}$ \label{line:dw_update_5}
\EndProcedure
\State 
\Procedure{Query}{} \Comment{Lemma \ref{lemma:distance_weight_query} }  
    \State \Return $s$ 
\EndProcedure
\State {\bf end data structure}
\end{algorithmic}
\end{algorithm}

Our algorithm starts with \textsc{Init}, which initializes its members $\{x_i\}_{i=1}^n, \{u_i\}_{i=1}^n, \{w_i\}_{i=1}^n, s$ and ADE as required. Next, when an online point $y$ comes, our algorithm calls \textsc{Update}$(y)$. Here the ADE will first output a series of approximate distances $\wt{d}_1,\wt{d}_2,\cdots,\wt{d}_n$. Next, our data structure will consider the approximate increments if matching $y$ to each offline point $x_i$. Next we will greedily select a node to match with $y$ and save its index as $i_0$. Next, we will update the corresponding $u_{i_0}$, $w_{i_0}$ and $s$. If matching value is needed, it will call \textsc{Query} to output current matching value $s$.

Algorithm \ref{alg:inner_product_weight} and Algorithm \ref{alg:inner_product_weight_v2} correspond to Theorem \ref{thm:inner_product_weight} and Theorem \ref{thm:inner_product_weight_v2} respectively, where the weight is described by the inner product. 
Their implementations are similar with Algorithm \ref{alg:distance_weight}. For Algorithm \ref{alg:inner_product_weight}, the adaptive distance estimation data structure ADE is replaced by our inner product estimation data structure to fit the inner product weight. Algorithm \ref{alg:inner_product_weight_v2} is more complex. It employs the Max-IP data structure (see Section \ref{thm:max-IP}) which is able to output the maximal inner product of a query point and given points on the unit sphere with high probability. To to apply this Max-IP data structure, our main idea is to add dimensions of both offline and online points to convert the process of finding the edge with the largest increment to the current matching into a single computing of inner product. This process can be achieved by calling \textsc{Query} using the points after asymmetric transformation (see Section \ref{section:other}). Through this idea, it achieves the sublinear running time of procedure \textsc{Update}.

\begin{algorithm}[!ht] \caption{Algorithm for online matching problem with inner product weight}\label{alg:inner_product_weight}
\begin{algorithmic}[1]
\State {\bf data structure} \textsc{InnerProductMatching} \Comment{Theorem \ref{thm:inner_product_weight} }
\State {\bf members}
\State \hspace{4mm} $x_1, x_2, \cdots x_n \in \R^{d}$ 
\State \hspace{4mm} $u_i$ (linklist or tree)
\State \hspace{4mm} $w_i$ (matching value on $x_i$)
\State \hspace{4mm} $s$ (matching value)
\State \hspace{4mm} \textsc{IPE} $\mathrm{Ipe}$
\State {\bf end members}
\State 
\Procedure{Init}{$x_1, \cdots, x_n, \epsilon, \delta$} \Comment{Lemma \ref{lemma:inner_product_weight_init} }
    \For {$i=1,2,\cdots,n$}
        \State $w_i \gets 0$
    \EndFor 
    \State $s \gets 0$
    \State $\mathrm{Ipe}$.\textsc{Init}($x_1, \cdots, x_n, \epsilon, \delta / n$)
\EndProcedure 
\State 
\Procedure{Update}{$y \in \R^d$} \Comment{Lemma \ref{lemma:inner_product_weight_update} }
    \State $\wt{w}_1,\wt{w}_2,\cdots,\wt{w}_n \gets \mathrm{Ipe}.\textsc{Query}(y)$ \label{line:ip_update_1}
    \State $i_0 \gets \arg\max_{i \in [n]}\{\wt{w}_i-w_i\}$ 
    \label{line:ip_update_2}
    \State $u_{i_0}.\textsc{Insert}(y)$ \label{line:ip_update_3}
    \State $s \gets s+\max\{w_{i_0},\wt{w}_{i_0}\}-w_{i_0}$ \label{line:ip_update_4}
    \State $w_{i_0} \gets \max\{w_{i_0},\wt{w}_{i_0}\}$ \label{line:ip_update_5}
\EndProcedure
\State 
\Procedure{Query}{} \Comment{Lemma \ref{lemma:inner_product_weight_query} }
    \State \Return $s$ 
\EndProcedure
\State {\bf end data structure}
\end{algorithmic}
\end{algorithm}

\begin{algorithm}[!ht] \caption{Better algorithm for online matching problem with inner product weight}\label{alg:inner_product_weight_v2}
\begin{algorithmic}[1]
\State {\bf data structure} \textsc{FasterInnerProductMatching}  
\Comment{Theorem \ref{thm:inner_product_weight_v2} }  
\State {\bf members}
\State \hspace{4mm} $X_1,X_2,\cdots,X_n \in \R^{d+1}$
\State \hspace{4mm} $w_1,w_2,\cdots,w_n\in\mathbb{R}$ (accumulated weight)
\State \hspace{4mm} $s\in\mathbb{R}$ (matching value)
\State \hspace{4mm} \textsc{MaxIP} $\mathrm{maxip}$
\State {\bf end members}
\State 
\Procedure{Init}{$x_1, \cdots, x_n, \epsilon, \tau, \delta$} \Comment{Lemma \ref{lemma:inner_product_weight_v2_init} } 
    \For {$i=1,2,\cdots,n$}
        \State $X_i \gets (x_i,0)$
    \EndFor 
    \State $s\gets 0$
    \State $\mathrm{maxip}$.\textsc{Init}($Q(X_1), \cdots, Q(X_n), 1-\epsilon, \tau/D, \delta / n$)
\EndProcedure 
\State 
\Procedure{Update}{$y \in \R^d$} \Comment{Lemma \ref{lemma:inner_product_weight_v2_update} }  
    \State $Y\gets (y,-1)$\label{line:ipv2_update_1}
    \If {$\mathrm{maxip}.\textsc{Query}(P(Y))$ succeeds}  \label{line:ipv2_if_begin}
        \State $(i,z) \gets \mathrm{maxip}.\textsc{Query}(P(Y))$ \label{line:ipv2_update_iz1}
    \Else 
        \State Select random $i$ from $U$, let $z=\Delta w_i$\label{line:ipv2_update_iz2}
    \EndIf \label{line:ipv2_if_end} 
    \State $u_{i}.\textsc{Insert}(y)$ \label{line:ipv2_update_insert}
    \If {$z>0$} \label{line:ipv2_if2_begin}
        \State $w_i \gets w_i + z$ 
        \State $s \gets s + z$
        \State $X_i\gets (x_i,w_i)$
        \State maxip.\textsc{Update}$(i,X_i)$
    \EndIf \label{line:ipv2_if2_end}
\EndProcedure
\State 
\Procedure{Query}{} \Comment{Lemma \ref{lemma:inner_product_weight_v2_query} }
    \State \Return $s$
\EndProcedure
\State {\bf end data structure}
\end{algorithmic}
\end{algorithm}

\section{More Details for Matching} \label{sec:missing_details}

In this section, we present more details on the definition of the online matching problem in Section \ref{sec:matching}. 
We also give the omitted algorithm of the greedy algorithm in Theorem \ref{thm:1/2-approximation}. We start by explaining our online matching problem to justify our motivation. Next, we introduce how our algorithm could achieve $1/2$-approximation as shown in Theorem \ref{thm:1/2-approximation}.

\subsection{Explanation of Online Matching Problem}

In the definition of the online matching problem (see Definition \ref{def:main_problem}), we require an online matching algorithm to take a point set $U$ (called offline point set) as input. Next for every coming point $v$ (called online point), we need to match it to a point in $U$. The weight of this edge is determined by weight function $w$.

We could allow one offline point $u\in U$ to match multiple online points due to the following reasons:
\begin{itemize}
    \item This problem is less important if the number of online points is larger than offline points.
    \item For the extreme situation: for every online point $v$ except from the first one, its best ``partner" (i.e. $\arg\max_{u\in U}(u,v)$) has always been matched by a former online point. It's easy to find that no $\Gamma$-approximation can be guaranteed in this case. The major reason is that this extreme case is possible for every online matching algorithm. So we give the offline points chances to ``regret". 
    In other words, we permit an offline point to discard its current partner and match with a new online point. 
    \item When the matching value is calculated, for every offline point, only the remaining partner will be taken into account.
\end{itemize}

\begin{algorithm}[!ht]\caption{Standard greedy online bipartite matching}\label{alg:standard_greedy}
\begin{algorithmic}[1]
\State {\bf data structure} \textsc{GreedyAlgorithm}
\State {\bf members}
\State \hspace{4mm} $x_1, x_2, \cdots x_n \in \R^{d}$
\State \hspace{4mm} $u_1,u_2,\cdots,u_n$ (Linklist or tree)
\State \hspace{4mm} $w_1,w_2,\cdots,w_n$ (accumulated weight of each offline point)
\State {\bf end members}
\Procedure{Init}{$x_1, \cdots, x_n$}
    \For{$i=1 \to n$}
        \State $x_i \gets x_i$
        \State Init $u_i$
        \State $w_i \gets 0$
    \EndFor
\EndProcedure
\Procedure{Update}{$y \in \R^d$}
    \State $w_{\max} \gets -1$  
    \State $i_{\max} \gets -1$  
    \For{$i = 1 \to n$}
        \If{$ \max\{w_{x_i,y}-w_i,0\} \geq w_{\max}$ } \Comment{$w_{x_i,y}$ is the weight of the edge $(x_i,y)$}
            \State $w_{\max} \gets \langle x_i, y \rangle$
            \State $i_{\max} \gets i$
        \EndIf
    \EndFor 
    \State $u_{i_{\max}}.\textsc{Insert}(y)$
    \State $w_{i_{\max}} \gets \max\{w_{i_{\max}},w_{x_{i_{\max}},y}\}$  
\EndProcedure
\State {\bf end data structure}
\end{algorithmic}
\end{algorithm}

\subsection{Greedy Algorithm}

In this section, we present a data structure in Algorithm \ref{alg:standard_greedy} that makes  $1/2$-approximation feasible.

This data structure contains the following components: (1) a series of $d$-dimensional vectors $x_1,x_2,\cdots,x_n$ used to save the coordinate of offline points, (2) a series of linklist or tree $u_1,u_2,\cdots,u_n$ where each $u_i$ is used to save the online nodes matched to the offline node $x_i$, (3) a series of real numbers $w_1,w_2,\cdots,w_n$ where each $w_i$ records the maximal edge value currently matched to $x_i$. Here we notice that linklists or trees $u_1,u_2,\cdots,u_n$ are not necessary, they are only for convenience for further uses. For example, users may want to quickly know how many online points an offline point $x_i$ is matched with so that we can quickly list them.

In the online bipartite matching phase, we start with operation \textsc{Init}, which stores offline points. Moreover, \textsc{Init} initializes the linklist or tree of each offline point. Furthermore \textsc{Init} initializes weights of each offline point to $0$. When an online point $y$ comes, we perform \textsc{Update}($y$). In this operation, $i_{\max}$ is used to record the optimal offline point and $w_{\max}$ is used to record the corresponding weight with $y$. \textsc{Update} simply judges each offline point to decide whether to change the values of $i_{\max}$ and $w_{\max}$, and after checking every offline node, it updates the corresponding $u_{i_{\max}}$ and $w_{i_{\max}}$.

Following theorem \ref{thm:1/2-approximation}, it is sufficient to show that we could achieve 1/2-approximation.

\section{Missing Proofs for Inner Product Weight}\label{sec:app_ip}

\subsection{Proof of Lemma~\ref{lemma:inner_product_weight_update}}

\begin{proof}
The running time of each line is stated as follows: 
\begin{itemize}
    \item Line \ref{line:ip_update_1} calls Ipe.\textsc{Query}, taking $\tilde{O}(\epsilon^{-2}D^2(n+d)\log(n/\delta))$ time by Theorem \ref{thm:IPE}.
    \item Line \ref{line:ip_update_2} computes $i_0$, taking $O(n)$ time.
    \item Line \ref{line:ip_update_3} updates $u_{i_0}$, taking at most $\tilde{O}(n)$ time.
    \item Line \ref{line:ip_update_4} updates $s$, taking $O(1)$ time.
    \item Line \ref{line:ip_update_5} updates $w_{i_0}$, taking $O(1)$ time.
\end{itemize}

To sum up, procedure \textsc{Update} takes $\tilde{O}(\epsilon^{-2}D^2(n+d)\log(n/\delta))$ time.
\end{proof}

\subsection{Proof of Lemma~\ref{lemma:inner_product_weight_query}}

\begin{proof}
Note that in \textsc{Update}, we call Ipe.\textsc{Query} to obtain a series of approximated weights related to an online point, match it to a node $x_{i_0}$ which increases the current matching value most, and store the updated matching value into $s$. Hence by Lemma \ref{lemma:main'}, suppose for every online point, when calling \textsc{Update}, the step Ipe.\textsc{Query} succeeds, that is, it succeeded in outputting a series of $\{\tilde{w}_i\}_{i=1}^n$ such that $w_i-\epsilon \le \tilde{w}_{i}\le w_i+\epsilon$, then \textsc{Query} will output a matching $\alg$ such that $\alg \ge \frac{1}{2}\opt - \frac{3}{2}n\epsilon$.

Consider the probability of success of each call of \textsc{Update}, it's at least $1 - \delta / n$ according to Theorem \ref{thm:IPE}, hence by union bound, the probability of succuss of all the \textsc{Update} is at least $1-\delta$, which means \textsc{Query} correctly outputs a matching $\alg$ such that $\alg \ge \frac{1}{2}\opt - \frac{3}{2}n\epsilon$ with probability at least $1-\delta$.
\end{proof}

\subsection{Proofs of Theorem~\ref{thm:inner_product_weight_v2}}

We prove the data structure (see Algorithm \ref{alg:inner_product_weight_v2}) satisfies the requirements Theorem \ref{thm:inner_product_weight_v2} by proving the following lemmas.  
\begin{lemma}\label{lemma:inner_product_weight_v2_init}
Procedure \textsc{Init} (Algorithm~\ref{alg:inner_product_weight_v2}) 
runs in $O(dn^{1+\rho} \log (n / \delta))$ time.
\end{lemma}
\begin{proof}
This is because maxip.\textsc{Init} takes $O(dn^{1+\rho} \log (n / \delta))$ time by Theorem \ref{thm:max-IP}, where $\rho$ and function $f$ follows from Theorem \ref{thm:max-IP}.
\end{proof}

\begin{lemma}\label{lemma:inner_product_weight_v2_update}
Procedure \textsc{Update} (Algorithm~\ref{alg:inner_product_weight_v2}) 
runs in $O(dn^{\rho} \log (n / \delta))$ time.
\end{lemma}
\begin{proof}

We explain the running time as follows:
\begin{itemize}
    \item Line \ref{line:ipv2_update_1} computes $Y$, taking $O(1)$ time.
    \item The if-statement from line \ref{line:ipv2_if_begin} to line \ref{line:ipv2_if_end} judges whether maxip.\textsc{Query}($P(Y)$) succeeds, taking $O(dn^{\rho} \log (n / \delta))$ time by Theorem \ref{thm:max-IP}.
    \item It also computes $i$ and $z$ in line \ref{line:ipv2_update_iz1} or line \ref{line:ipv2_update_iz2}, taking $O(1)$ time.
    \item Line \ref{line:ipv2_update_insert} inserts $y$ into $u_i$,  taking $\tilde{O}(d)$ time.
    \item The if-statement from line \ref{line:ipv2_if2_begin} to line \ref{line:ipv2_if2_end} updates $w_i, s, X_i$, taking $O(1)$ time, and calls maxip.\textsc{Update}$(i, X_i)$, taking $O(dn^{\rho} \log (n / \delta))$ time by Theorem \ref{thm:max-IP}. 
\end{itemize}
Thus the total time of procedure \textsc{Update} is $O(dn^{\rho} \log (n / \delta))$. 
\end{proof}
\begin{lemma}\label{lemma:inner_product_weight_v2_query}
Procedure \textsc{Query} (Algorithm~\ref{alg:inner_product_weight_v2}) 
correctly outputs a matching $\alg$ such that $\alg \ge \frac{1}{2}\min\{(1-\epsilon) \opt, \opt-n\tau\}$ with probability at least $1-\delta$, where $n$ is the number of online points.
\end{lemma}
\begin{proof}
Since $\langle Q(x), P(y) \rangle = \frac{1}{D} \langle x,y \rangle$, for an online point $v$, if there exists an offline point $u$ such that $\langle u,v\rangle \ge \tau$, then $\langle Q(u), P(y) \rangle \ge \tau/D$, then maxip.\textsc{Query} will give us a $u'$ with 
\begin{align*} 
\langle Q(u'), P(v)\rangle \ge (1-\epsilon) \langle Q(u), P(v)\rangle,
\end{align*}
which is equivalent to 
\begin{align*}
\langle Q(u'), P(v)\rangle \ge \langle Q(u), P(v) \rangle ;
\end{align*} 
if for all $u$, $\langle u,v \rangle < \tau$, then we randomly match $v$ to $u'$ with the increase of matching value $\Delta w(u',v) \ge 0 \ge \langle u,v \rangle -\tau$. Thus no matter how, we match $v$ to a $u$ such that
\begin{align*}
    w(u',v)\ge \min\{(1-\epsilon)\max_u\{w(u,v)\}, \max_u\{w(u,v)\}-\tau\}.
\end{align*}
Thus by Lemma \ref{lemma:greedy_robustness}, $\alg \ge \frac{1}{2}\min\{(1-\epsilon) \opt, \opt-n\tau\}$ holds as long as in each time of \textsc{Update}, maxip.\textsc{Query} succeeds. And by union bound, it happens with probability at least $1-\delta$, which finishes the proof.
\end{proof}







\end{document}